\documentclass[twocolumn,showpacs,preprintnumbers,amsmath,amssymb]{revtex4}

\usepackage{graphicx}

\newcommand{\bra}[1]{\langle #1|}
\newcommand{\ket}[1]{|#1\rangle}
\newcommand{\bracket}[1]{\langle#1\rangle}
\newcommand{\sub}[1]{_{\mathrm{#1}}}
\newcommand{\ii}{\mathrm{i}}
\newcommand{\ee}{\mathrm{e}}
\newcommand{\nn}{\nonumber \\}
\newcommand{\abs}[1]{\left\lvert#1\right\rvert}

\begin{document}

\title{Bloch sphere representation of three-vertex geometric phases}

\author{Shuhei Tamate}
\email{tamate@giga.kuee.kyoto-u.ac.jp}
\affiliation{%
Department of Electronic Science and Engineering, Kyoto University, Kyoto 615-8510, Japan
}%
\author{Kazuhisa Ogawa}
\affiliation{%
Department of Electronic Science and Engineering, Kyoto University, Kyoto 615-8510, Japan
}%
\author{Masao Kitano}
\affiliation{%
Department of Electronic Science and Engineering, Kyoto University, Kyoto 615-8510, Japan
}%

\date{\today}

\begin{abstract}
The properties of the geometric phases between three quantum states are investigated
in a high-dimensional Hilbert space using the Majorana representation of symmetric quantum states.
We found that the geometric phases between the three quantum states
in an $N$-state quantum system can be represented by
$N-1$ spherical triangles on the Bloch sphere.
The parameter dependence of the geometric phase was analyzed
based on this picture. We found that the geometric phase
exhibits rich nonlinear behavior in a high-dimensional
Hilbert space.
\end{abstract}

\maketitle

\section{Introduction}

The geometric phase was first discovered by Berry for
cyclic and adiabatic evolution in 1984 \cite{Berry1984}.
Subsequently, it was generalized to nonadiabatic evolution \cite{Aharonov1987}
and extended to noncyclic and nonunitary evolutions \cite{Samuel1988}.
The early studies by Pancharatnam \cite{Pancharatnam1956} on
the interference of polarized light were seminal for the generalization of geometric phases.
Pancharatnam proposed a method for comparing the phases
of two differently polarized light beams.
When two polarized light beams produce maximum constructive interference,
they are regarded as being {\it in phase}.
This means that two quantum states $\ket{\psi_1}$ and $\ket{\psi_2}$
are {\it in phase} when $\bracket{\psi_1|\psi_2}$ is real and positive \cite{Berry1987}.
Pancharatnam also pointed out that a geometric phase manifests itself
when three differently polarized states are successively compared 
using the {\it in-phase} relationship.

The generality of Pancharatnam's geometric phase is apparent in its kinematic definition.
It is defined without reference to a Hamiltonian or a dynamic evolution.
Mukunda and Simon further developed the kinematic approach for treating geometric phases \cite{Mukunda1993}.
In their studies, they focus on the Bargmann invariant \cite{Bargmann1964}.
For given three states $\ket{\psi_1}$, $\ket{\psi_2}$, and $\ket{\psi_3}$,
the three-vertex Bargmann invariant is defined as
$\bracket{\psi_1|\psi_3}\bracket{\psi_3|\psi_2}\bracket{\psi_2|\psi_1}$.
We refer to the argument of the Bargmann invariant as a
three-vertex geometric phase and define it as follows:
\begin{equation}
 \gamma(\psi_1,\psi_2,\psi_3) \equiv \arg\bracket{\psi_1|\psi_3}\bracket{\psi_3|\psi_2}\bracket{\psi_2|\psi_1}.  \label{eq:1}
\end{equation}
The above definition is equivalent to the quantum mechanical
expression for Pancharatnam's geometric phase.
The three-vertex geometric phase has been shown to equal
the geometric phase of the geodesic triangle that has the three states as vertices \cite{Samuel1988}.

As shown in \cite{Mukunda1993}, any pure-state geometric phase can be
constructed as a sum of three-vertex geometric phases.
Therefore, the three-vertex geometric phase serves as
a primitive building block for geometric phases.
An experiment has been performed to directly observe the three-vertex geometric phase
using a kinematic setup \cite{Kobayashi2010a}.

Three-vertex geometric phases are also important in
the field of quantum information
since they are closely related to distinguishability of
three quantum states \cite{Jozsa2000,Mitchison2004,Sugimoto2010}.

Despite its considerable importance,
little is known about the properties of the three-vertex geometric phase of
high-dimensional Hilbert spaces.
In this paper, we propose a way to represent three-vertex geometric phases for
$N$-state systems on the Bloch sphere and we then
investigate their characteristic behaviors
in high-dimensional Hilbert spaces.
For this purpose, we use the Majorana representation
of symmetric quantum states \cite{Majorana1932}.
Some theoretical studies have investigated geometric phases in
$N$-dimensional Hilbert space based on the Majorana representation \cite{Hannay1998a,Hannay1998}.

This paper is organized as follows.
In Sec.\,\ref{sec:Majorana}, we demonstrate how the three-vertex geometric
phase can be represented on the Bloch sphere with the Majorana representation.
In Sec.\,\ref{sec:quantum-eraser}, we review a measurement method for the
three-vertex geometric phase in a quantum eraser.
In Sec.\,\ref{sec:example}, we consider a three-state system as an example
and illustrate how the parameter dependence of the three-vertex geometric phase
can be visualized on the Bloch sphere.
We also show the nonlinear behavior of the geometric phase in
a high-dimensional Hilbert space.
Section~\ref{sec:conclusion} summarizes the findings of this study.

\section{Bloch sphere representation of three-vertex geometric phases} \label{sec:Majorana}

\begin{figure}
 \centering
 \includegraphics[width=140pt]{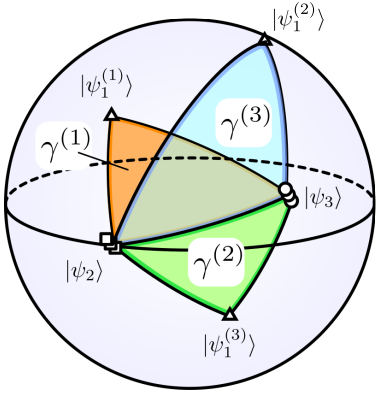}
 \caption{Bloch sphere representation of the three-vertex geometric phase of an $N$-state system. This figure shows $N=4$ case. The three-vertex geometric phase of a four-state system is represented as the sum of three geometric phases for a two-state system.}
 \label{fig:bloch}
\end{figure}

An $N$-dimensional Hilbert space $\mathcal{H}_N$ is isomorphic to
a symmetrized ($N-1$)-qubit Hilbert space:
\begin{equation}
 \mathcal{H}_N \simeq \mathcal{S} (\mathcal{H}_2{}^{\otimes {N-1}}),  \label{eq:2}
\end{equation}
where $\mathcal{S}$ denotes the projection operator onto the 
permutation-symmetric subspace.
We can construct an isomorphism map from $\mathcal{H}_N$
to $\mathcal{S}(\mathcal{H}_2{}^{\otimes {N-1}})$
by identifying the $i$th basis state of $\mathcal{H}_N$ to
the $(N-1)$-qubit symmetric Dicke state with $i-1$ excitations.
For example, for $N=3$, the three basis states $\ket{1}$, $\ket{2}$, and $\ket{3}$ in
$\mathcal{H}_3$ are identified with the states
$\ket{00}$, $(\ket{01} + \ket{10})/\sqrt{2}$, and $\ket{11}$
in $\mathcal{S}(\mathcal{H}_2\otimes \mathcal{H}_2)$.
In the following discussion, we identify the $N$-dimensional Hilbert space
with the symmetrized ($N-1$)-qubit Hilbert space.

According to Majorana \cite{Majorana1932}, an arbitrary symmetric state
$\ket{\varPsi}$ can be represented by the symmetrization of product states:
\begin{equation}
 \ket{\varPsi} = K\cdot \mathcal{S}(\ket{\psi^{(1)}}\cdots \ket{\psi^{(N-1)}}),  \label{eq:3}
\end{equation}
where $\ket{\psi^{(i)}}$ ($i=1\dots N-1$) are qubit states and $K$ is a normalization factor.
Equation~(\ref{eq:3}) is called the Majorana representation. We can depict the
state as an unordered set of $N-1$ points on the Bloch sphere;
these points are called Majorana points.

To characterize the geometric phase of the $N$-state system,
we first consider the following special set of three states:
\begin{align}
 \ket{\varPsi_1} &= K\cdot \mathcal{S}(\ket{\psi_1^{(1)}} \cdots \ket{\psi_1^{(N-1)}}),  \label{eq:4}\\
 \ket{\varPsi_2} &= \ket{\psi_2}\cdots\ket{\psi_2},  \label{eq:5}\\
 \ket{\varPsi_3} &= \ket{\psi_3}\cdots\ket{\psi_3}.  \label{eq:6}
\end{align}
The first state $\ket{\varPsi_1}$ is an arbitrary symmetric state and the other two, $\ket{\varPsi_2}$ and $\ket{\varPsi_3}$, are chosen from symmetric product states.
The three-vertex geometric phase of these three states
is expressed by:
\begin{align}
 \gamma(\varPsi_1, \varPsi_2, \varPsi_3)
 &= \arg \bracket{\varPsi_1|\varPsi_3}\bracket{\varPsi_3|\varPsi_2}\bracket{\varPsi_2|\varPsi_1} \nn
 &= \sum_{i=1}^{N-1} \arg \bracket{\psi_1^{(i)}|\psi_3}\bracket{\psi_3|\psi_2}\bracket{\psi_2|\psi_1^{(i)}} \nn
 &= \sum_{i=1}^{N-1} \gamma^{(i)},  \label{eq:7}
\end{align}
where $\gamma^{(i)} \equiv \gamma(\psi_1^{(i)}, \psi_2, \psi_3)$.
Equation~(\ref{eq:7}) implies that the geometric phase $\gamma(\varPsi_1, \varPsi_2, \varPsi_3)$
of an $N$-state system can be expressed as the sum of
the $N-1$ geometric phases $\gamma^{(i)}$ of a two-state system.
For the two-state system, the three-vertex geometric phase
$\gamma^{(i)}$ is proportional to the solid angle
$\varOmega^{(i)}$ of the geodesic triangle
with three vertices $\ket{\psi_1^{(i)}}$, $\ket{\psi_2}$, and $\ket{\psi_3}$;
that is, $\gamma^{(i)} = -\varOmega^{(i)}/2$ \cite{Pancharatnam1956}.
Therefore, we can regard the geometric phase
of the $N$-state system as $N-1$ spherical triangles
on the Bloch sphere, as shown in Fig.~\ref{fig:bloch}.

The validity of the above geometric description for
$\gamma(\varPsi_1, \varPsi_2, \varPsi_3)$
depends critically on the particular choice of the
three states.
However, an arbitrary set of three states in the $N$-state system can always
be transformed into the set of forms shown in Eqs.~(\ref{eq:4})--(\ref{eq:6})
by a unitary transformation.
We assume that $\ket{\varPhi_1}$, $\ket{\varPhi_2}$, and $\ket{\varPhi_3}$ are
an arbitrary set of three states.
There exists a pair of symmetric product states, $\ket{\varPsi_2}$ and $\ket{\varPsi_3}$,
such that $\bracket{\varPsi_2|\varPsi_3} = \bracket{\varPhi_2|\varPhi_3}$
since the inner product of a pair of symmetric product states
can attain an arbitrary complex number with a modulus less than or equal to 1.
Then, we can always find a unitary operator $\hat{U}$ such that
$\ket{\varPsi_2} = \hat{U}\ket{\varPhi_2}$, $\ket{\varPsi_3} = \hat{U}\ket{\varPhi_3}$.
Finally, $\ket{\varPsi_1}$ is chosen to satisfy $\ket{\varPsi_1} = \hat{U} \ket{\varPhi_1}$.

As geometric phases are invariant under unitary transformations,
\begin{equation}
 \gamma(\varPhi_1, \varPhi_2, \varPhi_3) = \gamma(\varPsi_1, \varPsi_2, \varPsi_3).  \label{eq:8}
\end{equation}
Therefore, we can represent an arbitrary
three-vertex geometric phase on the Bloch sphere by applying
a proper unitary transformation.

\section{Geometric phase in quantum erasers} \label{sec:quantum-eraser}

\begin{figure}
 \centering
 \includegraphics[width=210pt]{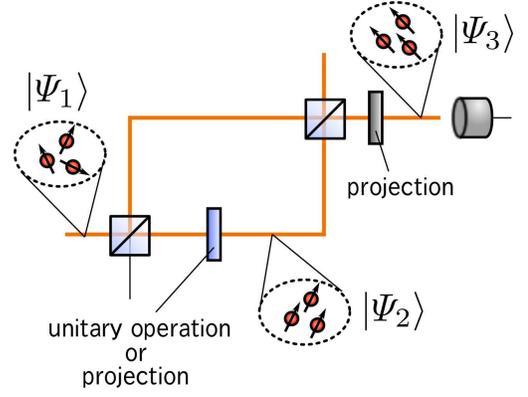}
 \caption{Scheme for measuring three-vertex geometric phases in
 a quantum eraser.
 We first prepare an internal state to be $\ket{\varPsi_1}$ and split the beam into
 two. The internal state of the lower beam is then transformed
 into state $\ket{\varPsi_2}$. Finally, the two beams are combined and
 the interference pattern is observed with or without a projector to the
 internal state. We can extract the three-vertex
 geometric phase from the shift of the interference fringes.}
 \label{fig:interferometer}
\end{figure}

In this section, we briefly describe a method for measuring
the three-vertex geometric phases in a quantum eraser \cite{Tamate2009a,Kobayashi2011}.

We consider a particle interferometer with internal degrees of freedom,
as shown in Fig.~\ref{fig:interferometer}.
Assuming that the particle has an initial state of $\ket{\varPsi_1}$,
the state of the composite system after the first beam splitter can be written as
\begin{equation}
 \ket{\varPhi\sub{i}}\rangle = \ket{\varPsi_1}\ket{+},  \label{eq:9}
\end{equation}
where $\ket{+} = (\ket{0} + \ket{1})/\sqrt{2}$ is a
superposition of the upper path state $\ket{0}$ and the lower path state $\ket{1}$.
In the lower path, the internal state is transformed to a state $\ket{\varPsi_2}$
and we have the following intermediate state of the composite system:
\begin{equation}
 \ket{\varPhi\sub{m}}\rangle = \frac{1}{\sqrt{2}}(\ket{\varPsi_1}\ket{0} + \ket{\varPsi_2}\ket{1}).  \label{eq:10}
\end{equation}
We then project the internal state onto $\ket{\varPsi_3}$ and obtain the final state:
\begin{equation}
 \ket{\varPhi\sub{f}}\rangle = K\ket{\varPsi_3}(\bracket{\varPsi_3|\varPsi_1}\ket{0} + \bracket{\varPsi_3|\varPsi_2}\ket{1}),  \label{eq:11}
\end{equation}
where $K = (\abs{\bracket{\varPsi_3|\varPsi_1}}^2 + \abs{\bracket{\varPsi_3|\varPsi_2}}^2)^{-1/2}$ is the normalization factor.
The final measurement for the path state is represented by the projector
\begin{align}
 \hat{P}(\delta) &= \hat{1}\otimes \ket{\delta}\bra{\delta},  \label{eq:12}\\
 \ket{\delta} &= \frac{1}{\sqrt{2}}(\ket{0} + \ee^{\ii \delta}\ket{1}),  \label{eq:13}
\end{align}
where $\delta$ is the phase difference between the upper and lower paths.
The output probability $P(\delta)$ is given by
\begin{align}
 P(\delta) &= \langle\langle\varPhi\sub{f}|\hat{P}(\delta)|\varPhi\sub{f}\rangle\rangle \nn
 &= \frac{1}{2} ( 1 + V \cos \varphi ),  \label{eq:14}
\end{align}
where
\begin{align}
 V &= \frac{2\abs{\bracket{\varPsi_1|\varPsi_3}\bracket{\varPsi_3|\varPsi_2}}}%
 {\abs{\bracket{\varPsi_3|\varPsi_1}}^2 + \abs{\bracket{\varPsi_3|\varPsi_2}}^2},  \label{eq:15}\\
 \varphi &= \arg\bracket{\varPsi_1|\varPsi_3}\bracket{\varPsi_3|\varPsi_2} - \delta.  \label{eq:16}
\end{align}
We change the parameter $\delta$ and thereby measure the interference fringes.
We can extract the constructive interference point $\delta\sub{f}$
by setting $\varphi = 0$ as
\begin{equation}
 \delta\sub{f} = \arg\bracket{\varPsi_1|\varPsi_3}\bracket{\varPsi_3|\varPsi_2}.  \label{eq:17}
\end{equation}
Similarly, we can obtain the constructive interference point $\delta\sub{m}$
for the case without final projection of the internal state as
\begin{equation}
 \delta\sub{m} = \arg\bracket{\varPsi_1|\varPsi_2}.  \label{eq:18}
\end{equation}
As a result, we can extract the geometric phase $\gamma$ of the three states
as the difference of $\delta\sub{f}$ and $\delta\sub{m}$:
\begin{equation}
 \gamma = \delta\sub{f} - \delta\sub{m}
  = \arg\bracket{\varPsi_1|\varPsi_3}\bracket{\varPsi_3|\varPsi_2}\bracket{\varPsi_2|\varPsi_1}.  \label{eq:19}
\end{equation}
This method enables us to measure the three-vertex geometric phase
using only state preparation, projection, and interference.

\section{Parameter dependence of the three-vertex geometric phase} \label{sec:example}

\begin{figure}
 \centering
 \includegraphics[width=150pt]{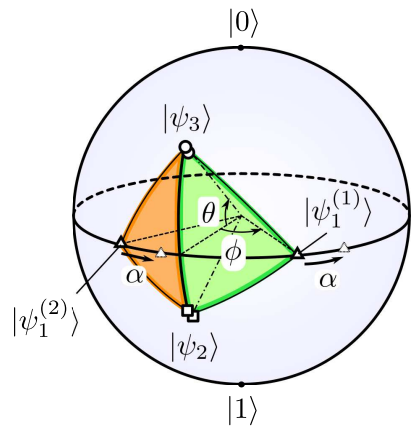}
 \caption{Variation of the Majorana points. The parameter $\theta$ represents the half
 angle between the states $\ket{\psi_2}$ and $\ket{\psi_3}$. The parameter $\phi$
 represents the half angle between the states $\ket{\psi_1^{(1)}(\alpha)}$ and
 $\ket{\psi_1^{(2)}(\alpha)}$. The two states $\ket{\psi_1^{(1)}(\alpha)}$
 and $\ket{\psi_1^{(2)}(\alpha)}$ are rotated on the Bloch sphere
 in accordance with the parameter $\alpha$.}
 \label{fig:bloch_evolution}
\end{figure}

To investigate the characteristic behavior of the three-vertex geometric phase, we consider a three-state system as an example.

In a three-state system, the Majorana representation that corresponds to Eqs.~(\ref{eq:4})--(\ref{eq:6}) is given by
\begin{align}
 \ket{\varPsi_1} &= K(\ket{\psi_1^{(1)}}\ket{\psi_1^{(2)}} + \ket{\psi_1^{(2)}}\ket{\psi_1^{(1)}}),  \label{eq:20}\\
 \ket{\varPsi_2} &= \ket{\psi_2}\ket{\psi_2},  \label{eq:21}\\
 \ket{\varPsi_3} &= \ket{\psi_3}\ket{\psi_3}.  \label{eq:22}
\end{align}
The normalization factor is calculated as $K = [2(1+|\bracket{\psi_1^{(1)}|\psi_1^{(2)}}|^2)]^{-1/2}$.
We consider the case where each qubit state is given as follows:
\begin{align}
 \ket{\psi_1^{(1)}} &= \frac{1}{\sqrt{2}}(\ee^{-\ii\phi/2}\ket{0} + \ee^{\ii\phi/2}\ket{1}),  \label{eq:23}\\
 \ket{\psi_1^{(2)}} &= \frac{1}{\sqrt{2}}(\ee^{\ii\phi/2}\ket{0} + \ee^{-\ii\phi/2}\ket{1}),  \label{eq:24}\\
 \ket{\psi_2} &= \cos(\theta/2)\ket{+} - \sin(\theta/2)\ket{-},  \label{eq:25}\\
 \ket{\psi_3} &= \cos(\theta/2)\ket{+} + \sin(\theta/2)\ket{-}.  \label{eq:26}
\end{align}
where $\ket{0}$ and $\ket{1}$ are the orthonormal basis states
and $\ket{\pm} = (\ket{0} \pm \ket{1})/\sqrt{2}$.
The Majorana points for Eqs.~(\ref{eq:23})--(\ref{eq:26}) are depicted in Fig.~\ref{fig:bloch_evolution}.

We introduce a parameter $\alpha$ for the state $\ket{\varPsi_1}$
by applying a unitary operator $\hat{U}(\alpha)$ to each qubit:
\begin{align}
 \ket{\varPsi_1(\alpha)} &= \hat{U}(\alpha)\otimes\hat{U}(\alpha)\ket{\varPsi_1},  \label{eq:27}\\
 \hat{U}(\alpha) &= \exp(-\ii \alpha \hat{\sigma}_z/2),  \label{eq:28}
\end{align}
where $\hat{\sigma}_z = \ket{0}\bra{0} - \ket{1}\bra{1}$.
Some calculation leads to
\begin{align}
 \ket{\varPsi_1(\alpha)} &= K(\ket{\psi_1^{(1)}(\alpha)}\ket{\psi_1^{(2)}(\alpha)} + \ket{\psi_1^{(2)}(\alpha)}\ket{\psi_1^{(1)}(\alpha)}),  \label{eq:29}\\
 \ket{\psi_1^{(1)}(\alpha)}
&= \frac{1}{\sqrt{2}}(\ee^{-\ii(\phi + \alpha)/2}\ket{0} + \ee^{\ii(\phi + \alpha)/2}\ket{1}),  \label{eq:30}\\
 \ket{\psi_1^{(2)}(\alpha)} 
&= \frac{1}{\sqrt{2}}(\ee^{\ii(\phi - \alpha)/2}\ket{0} + \ee^{-\ii(\phi - \alpha)/2}\ket{1}).  \label{eq:31}
\end{align}
As shown in Eq.~(\ref{eq:7}), the three-vertex geometric phase $\gamma$ for
the three-state system ($N=3$) is the sum of the geometric phases for two qubits:
\begin{equation}
 \gamma (\alpha) = \gamma^{(1)}(\alpha) + \gamma^{(2)}(\alpha).  \label{eq:32}
\end{equation}
Assuming $-\pi/2 < \theta < \pi/2$,
the geometric phase of each qubit can be expressed as
\begin{align}
 \gamma^{(1)}(\alpha) &= 2 \tan^{-1}\left(\tan \frac{\theta}{2} \tan \frac{\phi + \alpha}{2}\right),  \label{eq:33}\\
 \gamma^{(2)}(\alpha) &= - 2 \tan^{-1}\left(\tan \frac{\theta}{2} \tan\frac{\phi - \alpha}{2}\right).  \label{eq:34}
\end{align}

\begin{figure}[t]
 \centering
 \includegraphics[width=200pt]{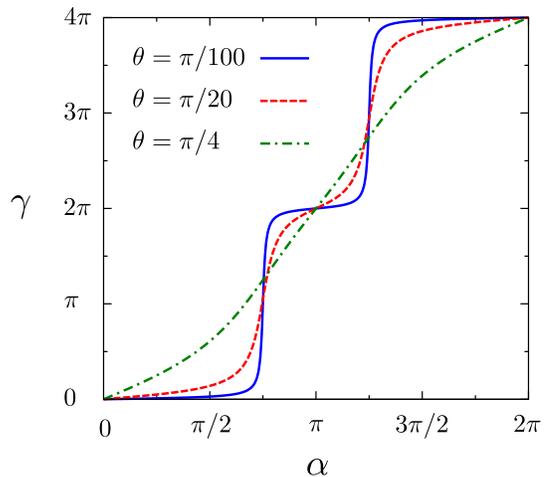}
 \caption{Variation of three-vertex geometric phase.
 The angle $\phi$ is kept constant at $\pi/4$. There are two singular points
 at $\alpha = 3\pi/4$ and $5\pi/4$. As the angle $\theta$ decreases, the
 slopes at these singular points increases.}
 \label{fig:gp3}
\end{figure}

Figure~\ref{fig:gp3} shows the variation of the three-vertex geometric phase.
The parameter $\phi$ is kept constant at $\pi/4$ while
the geometric phase $\gamma$ is varied between $0$ and $4\pi$.
The three-vertex geometric phase varies extremely nonlinearly at points 
$\alpha = 3\pi/4$ and $5\pi/4$.
This is closely related to the nonlinear behavior of the three-vertex geometric phases
of two-state systems \cite{Schmitzer1993,Li1999,Bhandari1991,Tamate2009a,Kobayashi2011}.
The nonlinearity in the two-state systems is due to the geometry
of the Bloch sphere and it originates from
the drastic change in the geodesic arcs surrounding the spherical triangle.
The Majorana representation implies that we can understand
the nonlinear behavior of the geometric phase of three-state systems
in a similar manner to that of two-state systems.
The three-state system has two corresponding spherical triangles.
The variation of the three-vertex geometric phase can be interpreted
as the nonlinear behavior of these triangles.
When $\alpha = 3\pi/4$ and $5\pi/4$, the points corresponding to $\ket{\psi_1^{(1)}(\alpha)}$
and $\ket{\psi_1^{(2)}(\alpha)}$ move around the back side of the Bloch sphere and the areas of the spherical triangles change rapidly.

While the behavior of the geometric phase in this example is
related to the geometric phase of two-state systems,
it differs qualitatively from the geometric phase of
two identical qubit states \cite{Klyshko1989,Kobayashi2011a},
which is simply twice the geometric phase of a one qubit state.
It is possible to flexibly control the geometric phase of a three-state system
by considering the arrangement of Majorana points.

These results can be extended to $N$-state systems in a straightforward manner.
The variation of the three-vertex geometric phase in an $N$-state system
can be decomposed into the contributions of the $N-1$ three-vertex geometric phases
in a two-state system.
Although three-vertex geometric phases of $N$-state systems
have more complex and interesting behaviors than those of two state systems,
they can be understood simply as a sum of the $N-1$ three-vertex geometric phases
in a two-state system.

Finally, we note that a three-state system can be realized by the polarization of
two photons in the same spatiotemporal mode, a biphotonic qutrit \cite{Bogdanov2004}.
It has been demonstrated that an arbitrary qutrit state can be well prepared
in a biphotonic system \cite{Krivitski2005}.
Therefore, the nonlinear behavior described in this section can be observed
by directly applying the measurement method in Sec.~\ref{sec:quantum-eraser} to
a biphotonic qutrit system, which is similar to the setup used in \cite{Kobayashi2011a}.

\section{Summary} \label{sec:conclusion}

We presented a way to represent three-vertex geometric phases
on the Bloch sphere. Our method is based on the Majorana representation
and the three-vertex geometric phase is represented as a set
of $N-1$ spherical triangles connecting the corresponding Majorana points.
We considered a three-state system as an example and
examined the parameter dependence of the three-vertex geometric phase.
We showed that the three-vertex geometric phase exhibits interesting
nonlinear behavior in a high-dimensional Hilbert space.
The characteristic behavior of
the three-vertex geometric phase in the $N$-state system can
be well understood by decomposing it into the
$N-1$ three-vertex geometric phases of a two-state system.

\acknowledgments
This research is supported by the Global COE Program
``Photonics and Electronics Science and Engineering'' at Kyoto University.
One of the authors (S.T.) is supported by a JSPS Research Fellowship for Young Scientists (No. 224850).

\end{document}